\documentclass[10pt]{article}
\usepackage{amsmath}
\usepackage{amsthm}
\usepackage{amssymb}
\usepackage[dvips]{graphicx}

\makeatletter
\@addtoreset{equation}{section}
\renewcommand{\theequation}{\thesection.\@arabic\c@equation}
\makeatother

\newtheorem{theorem}{Theorem}

\newtheorem{proposition}{Proposition}
\newtheorem{lemma}{Lemma}

\begin{document}
\title{Bifurcation diagrams of the Kowalevski top \\
in two constant fields\thanks{REGULAR AND CHAOTIC DYNAMICS, V. 10, N
4, 2005, pp. 381-398, DOI: 10.1070/RD2005v010n04ABEH000321}}
\author{Mikhail P. Kharlamov}
\date{}


\maketitle

\begin{abstract}
The Kowalevski top in two constant fields is known as the unique
profound example of an integrable Hamiltonian system with three degrees
of freedom not reducible to a family of systems in fewer dimensions.
As the first approach to topological analysis of this system we find
the critical set of the integral map; this set consists of the
trajectories with number of frequencies less than three. We obtain
the equations of the bifurcation diagram in ${\bf R}^3 $.
A correspondence to the Appelrot classes in the classical Kowalevski
problem is established. The admissible regions for the values of the
first integrals are found in the form of some inequalities of
general character and boundary conditions for the induced diagrams
on  energy levels.
\end{abstract}

\section{Introduction}\label{sec1}
During the last 20 years the integrable case of S.~Kowalevski
\cite{b1} has received several generalizations. Among them a special
place is given to the case \cite{b2} of rotation about a fixed point
of a heavy electrically charged gyrostat in gravitational and
electric fields of force. For a rigid body without gyrostatic
effects the corresponding equations were first considered in \cite{b3}
and interpreted as the equations of motion of a massive magnet subject to the
gravity force and constant magnetic force fields. The mathematical
model of superposition of such fields is referred to as {\it two
constant fields} \cite{b4}.

The case \cite{b2} does not have any explicit groups of symmetry and
therefore provides an illustration of a physically realizable system
with three degrees of freedom not admitting any obvious reduction to
a family of systems with two degrees of freedom. The phase topology of
irreducible systems has not been studied yet. The theory of
$n$-dimensional integrable systems originated in \cite{b5} has not been
further developed due to the absence, at that moment, of non-trivial
natural examples.

The result \cite{b2} succeeded some previous publications dealing
with rigid bodies and gyrostats satisfying the conditions of
Kowalevski type: I.V.~Komarov \cite{b6} has proved the complete
integrability of the Kowalevski gyrostat in gravity force field by
finding the first generalization of the Kowalevski integral $K$; the
corresponding integral for the rigid body in two constant fields was
pointed out by O.I.~Bogoyavlensky \cite{b3}; this integral was
upgraded to the case of gyrostat by H.~Yehia \cite{b8}. Yet the
analog of the Kowalevski case for two constant fields had not been
considered integrable until A.G.~Reyman and M.A.~Semenov-Tian-Shansky
\cite{b2} found the Lax representation with spectral parameter; this
immediately led to the new integral generalizing the square of
momentum integral for axially symmetric force fields.

Later, in a joint publication with A.I.~Bobenko \cite{b4}, the authors
of \cite{b2} presented algebraic foundations for the integrability
of multidimensional Kowalevski gyrostats and described a viable
way of explicit integration using finite-band technique. For two
constant fields this integration was never fulfilled.

This paper starts the investigation of three-dimensional phase
topology of a rigid body of Kowalevski type in two constant fields.

\section{Preliminaries}\label{sec2}
Consider a rigid body with fixed point $O$. Choose a trihedral at
$O$ rotating along with the body and refer to it all vector and
tensor objects. Denote by ${\bf e}_1{\bf e}_2{\bf e}_3$ the
canonical unit basis in ${\bf R}^3$; then the moving trihedral
itself is represented as $O{\bf e}_1{\bf e}_2{\bf e}_3$.

Constant field is a force field inducing the rotating moment about
$O$ of the form
\begin{equation}\label{q2_2}
{\bf r}\times{\boldsymbol \alpha}
\end{equation}
with constant vector ${\bf r}$ and with ${\boldsymbol \alpha}$
corresponding to some physical vector fixed in inertial space;
${\bf r}$ points from $O$ to the center of application of the field,
${\boldsymbol \alpha}$ is the field's intensity.

For two constant fields the rotating moment is ${\bf
r}_1\times{\boldsymbol \alpha}+{\bf r}_2\times{\boldsymbol \beta}$.
It can be represented as (\ref{q2_2}) if either ${\bf r}_1\times{\bf
r}_2=0$ or ${\boldsymbol \alpha}\times{\boldsymbol \beta}=0$. In the
sequel we suppose that
\begin{equation}\label{q2_3}
{\bf r}_1\times{\bf r}_2 \ne 0, \quad {\boldsymbol
\alpha}\times{\boldsymbol \beta} \ne 0.
\end{equation}

Two constant fields satisfying (\ref{q2_3}) are said to be {\it
independent}.

Introduce some notation.

Let $L(n,k)$ be the space of $n \times k$-matrices. Put
$L(k)=L(k,k)$.

Identify ${\bf R}^6={\bf R}^3 \times {\bf R}^3$ with  $L(3,2)$ by
the isomorphism $j$ that joins two columns
$$
A=j({\bf a}_1 ,{\bf a}_2 ) = \left\| {{\bf a}_1 \;{\bf a}_2 }
\right\| \in L(3,2), \qquad {\bf a}_1 ,{\bf a}_2 \in {\bf R}^3.
$$

For the inverse map, we write
$$
j^{ - 1} (A) = ({\bf c}_1 (A),{\bf c}_2 (A)) \in {\bf R}^3 \times
{\bf R}^3, \qquad A \in L(3,2).
$$

If $A,B \in L(3,2)$, ${\bf a} \in {\bf R}^3 $, by definition, put
\begin{equation} \label{q2_4}
A \times B = \sum\limits_{i = 1}^2 {{\bf c}_i (A) \times {\bf c}_i
(B) \in {\bf R}^3 ;}\quad {\bf a} \times A = j({\bf a} \times {\bf
c}_1 (A),{\bf a} \times {\bf c}_2 (A)) \in L(3,2).
\end{equation}

\begin{lemma}\label{lem1} Let $\Lambda  \in SO(3)$, $D \in GL(2,{\bf R})$, ${\bf
a} \in {\bf R}^3 $, $A,B \in L(3,2)$. Then
\begin{equation}\label{q2_5} \notag
\begin{array}{ll}
\Lambda (A \times B) = (\Lambda A) \times (\Lambda B); &
(AD^{-1})\times (BD^T) = A \times B; \\
\Lambda ({\bf a} \times A) = (\Lambda {\bf a}) \times (\Lambda A); &
{\bf a} \times (AD) = ({\bf a} \times A)D.
\end{array}
\end{equation}
\end{lemma}

The proof is by direct calculation.

Denote by ${\bf I}$ the inertia tensor of the body at $O$ and by
${\boldsymbol \omega }$ the angular velocity. Using the notation
(\ref{q2_4}) we write the Euler - Poisson equations of motion in the
form
\begin{equation}\label{q2_6}
{\bf I}{\boldsymbol \omega }^{\boldsymbol \cdot} = {\bf
I}{\boldsymbol \omega } \times {\boldsymbol \omega } + A \times
U,\qquad U^{\boldsymbol \cdot} =  - {\boldsymbol \omega } \times U.
\end{equation}
Here $A=j({\bf r}_1,{\bf r}_2)$ is a constant matrix,
$U=j({\boldsymbol \alpha},{\boldsymbol \beta})$. The phase space of
(\ref{q2_6}) is $\{({\boldsymbol \omega},U)\}={\bf R}^3\times
L(3,2)$.

In fact, $U$ in (\ref{q2_6}) is restricted by geometric integrals;
that is, for some constant symmetric $C \in L(2)$
\begin{equation}\label{q2_7}
U^T U = C.
\end{equation}

Let ${\mathcal O}$ represent the set (\ref{q2_7}) in $L(3,2)$. In order
to emphasize the  $C$-dependence, we write ${\mathcal O}={\mathcal O}(C)$.

Let $S=({\bf I},A,C)$. Denote by $X_S$ the vector field on ${\bf
R}^3 \times {\mathcal O}(C)$ corresponding to the system
(\ref{q2_6}).

Associate to $\Lambda \in SO(3)$, $D \in GL(2,{\bf R})$ the linear
automorphisms $\Psi(\Lambda,D)$ and $\psi(\Lambda,D)$ of ${\bf R}^3
\times L(3,2)$ and $L(3) \times L(3,2) \times L(2)$
\begin{equation}\label{q2_8}
\begin{array}{l}
\Psi(\Lambda,D)({\boldsymbol \omega},U)=(\Lambda {\boldsymbol
\omega},\Lambda U
D^T), \\
\psi(\Lambda,D)({\bf I},A,C)=(\Lambda {\bf I} \Lambda^T, \Lambda A
D^{-1}, D C D^T).
\end{array}
\end{equation}

It is easy to see that (\ref{q2_7}) and (\ref{q2_8}) imply
$\Psi(\Lambda,D)({\bf R}^3 \times {\mathcal O}(C))={\bf R}^3 \times
{\mathcal O}(D C D^T)$. Using Lemma~\ref{lem1} we obtain the
following statement.

\begin{lemma}\label{lem2}
For each $(\Lambda,D) \in SO(3)\times GL(2,{\bf R})$, we have
$$
\Psi(\Lambda,D)_*(X_S(v))=X_{\psi(\Lambda,D)(S)}(\Psi(\Lambda,D)(v)),
\qquad v \in {\bf R}^3 \times {\mathcal O}(C).
$$
\end{lemma}

Thus, any two problems of rigid body dynamics in two constant fields
(for short, RBD-problems) determined by the sets of parameters $S$
and $\psi(\Lambda,D)(S)$ are completely equivalent.

Let us call an RBD-problem {\it canonical} if the centers of
application of forces lie on the first two axes of the moving
trihedral at unit distance from $O$ and the intensities of the
forces are orthogonal to each other.

\begin{proposition} \label{prop1}
For each RBD-problem with independent forces there exists an equivalent canonical problem. Moreover, in both equivalent problems the centers of
application of forces belong to the same  plane in the body containing the
fixed point.
\end{proposition}
\begin{proof} Let the RBD-problem determined by the set of
parameters $S=({\bf I},A,C)$ satisfy (\ref{q2_3}). This means that
the symmetric matrices $A_*=(A^TA)^{-1}$ and $C$ are positively
definite.

According to the well-known fact from linear algebra, $A_*$ and $C$ can
be reduced, respectively, to the identity matrix and to a diagonal
matrix via the same conjugation operator
\begin{equation}\label{q2_9} \notag
DA_*D^T=E,\quad DCD^T= \mathop{\rm diag}\nolimits \{a^2,b^2\},\qquad
D \in GL(2,{\bf R}),\;a,b \in {\bf R}_+.
\end{equation}

Then ${\bf c}_1(AD^{-1})$ and ${\bf c}_2(AD^{-1})$ form an
orthonormal pair in ${\bf R}^3$. There exists $\Lambda \in SO(3)$
such that $\Lambda{\bf c}_i(AD^{-1})={\bf e}_i \,(i=1,2)$. The first
statement is obtained by applying Lemma \ref{lem2} with the previously chosen
$\Lambda,D$ to the initial vector field $X_S$.

To finish the proof, notice that the transformation $A \mapsto
AD^{-1}$ preserves the plane spanning ${\bf c}_1(A)$, ${\bf
c}_2(A)$. The matrix $\Lambda$ in (\ref{q2_8}) represents the
change of the moving trihedral. Therefore, if ${\bf a}\in {\bf
R}^3$ represents some physical vector in the initial problem, then
$\Lambda{\bf a}$ is the same vector with respect to the body in the
equivalent problem.
\end{proof}

{\bf Remark}. The fact that any RBD-problem can be reduced to the
problem with one of the pairs ${\bf r}_1,{\bf r}_2$ or ${\boldsymbol
\alpha}, {\boldsymbol \beta}$ orthonormal is known from \cite{b9}.
Simultaneous orthogonalization of both pairs crucially simplifies
calculations below.

It follows from Proposition~\ref{prop1} that, without loss of
generality, for independent forces we may suppose
\begin{gather}
{\bf r}_1={\bf e}_1,\quad {\bf r}_2={\bf e}_2, \label{q2_10} \\
{\boldsymbol \alpha} \cdot {\boldsymbol \alpha}=a^2, \;{\boldsymbol
\beta} \cdot {\boldsymbol \beta}=b^2, \;{\boldsymbol \alpha} \cdot
{\boldsymbol \beta}=0. \label{q2_11}
\end{gather}

Change, if necessary, the order of ${\bf e}_1, {\bf e}_2$ (with
simultaneous change of the direction of ${\bf e}_3$) to obtain $a
\geqslant b > 0$.

Consider a dynamically symmetric top in two constant fields with the
centers of application of forces in the equatorial plane of its
inertia ellipsoid. Choose a moving trihedral such that $O{\bf e}_3$
is the symmetry axis. Then the inertia tensor ${\bf I}$ becomes
diagonal. Let $a=b$. For any $\Theta \in SO(2)$ denote by ${\hat
\Theta} \in SO(3)$ the corresponding rotation of ${\bf R}^3$ about
$O{\bf e}_3$. Take in  (\ref{q2_8}) $\Lambda={\hat \Theta}$,
$D=\Theta$. Under the conditions (\ref{q2_10}), (\ref{q2_11}) $\psi=
\textrm{Id}$ and $\Psi$ becomes the symmetry group. The system
(\ref{q2_6}) has the cyclic integral $ {\bf I}{\boldsymbol
\omega}\cdot (a^2{\bf e}_3 - {\boldsymbol \alpha}\times {\boldsymbol
\beta})$ pointed out in \cite{b8} for the analog of the Kowalevski
case. Therefore it is possible to reduce such an RBD-problem to a
family of systems with two degrees of freedom.

Let us call an RBD-problem {\it irreducible} if for its canonical
representation (\ref{q2_10}), (\ref{q2_11}) the following inequality holds
\begin{equation}\label{q2_12}
a>b>0.
\end{equation}

The following statements are needed in the future; they also reveal
some features of a wide class of RBD-problems.
\begin{lemma} \label{lem3}
In an irreducible RBD-problem, the body has exactly four
equilibria.
\end{lemma}

\begin{proof}
The set of singular points of (\ref{q2_6}) is defined by $
{\boldsymbol \omega}=0,\;A \times U=0$. For the equivalent canonical
problem with (\ref{q2_10})
\begin{equation}\label{q2_13}
{\bf e}_1 \times {\boldsymbol \alpha}+{\bf e}_2 \times {\boldsymbol
\beta}=0.
\end{equation}

Then the four vectors in (\ref{q2_13}) are parallel to the same
plane and $\left| {\bf e}_1 \times {\boldsymbol \alpha}
\right|=\left| {\bf e}_2 \times {\boldsymbol \beta} \right|$. With
(\ref{q2_11}), (\ref{q2_12}) this equality yields
\begin{equation}\label{q2_14}
{\boldsymbol \alpha}=\pm a{\bf e}_1, \; {\boldsymbol \beta}=\pm b
{\bf e}_2.
\end{equation}

From mechanical point of view, the result is absolutely clear: none
of the orthogonal forces with unequal intensities and
"orthonormal"\, centers of application can produce a non-zero moment
at an equilibrium.
\end{proof}

\begin{lemma} \label{lem4}
Let an irreducible RBD-problem in its canonical form have the
diagonal inertia tensor ${\bf I}$. Then the body has the following
families of periodic motions of pendulum type
\begin{gather}
\begin{array}{c}
{\boldsymbol \alpha } \equiv  \pm a{\bf e}_1,\quad {\boldsymbol
\omega } = \varphi ^ {\boldsymbol \cdot}  {\bf e}_1 , \quad
{\boldsymbol \beta } = b({\bf e}_2 \cos
\varphi - {\bf e}_3 \sin \varphi ), \\
2\varphi ^{ {\boldsymbol \cdot}  {\boldsymbol \cdot} }  =  - b\sin
\varphi ;
\end{array} \label{q2_15}\\
\begin{array}{c}
{\boldsymbol \beta } \equiv  \pm b{\bf e}_2 , \quad {\boldsymbol
\omega } = \varphi ^ {\boldsymbol \cdot}  {\bf e}_2 , \quad
{\boldsymbol \alpha } = a({\bf
e}_1 \cos \varphi + {\bf e}_3 \sin \varphi ), \\
2\varphi ^{ {\boldsymbol \cdot}  {\boldsymbol \cdot} }  =  - a\sin
\varphi;
\end{array} \label{q2_16} \\
\begin{array}{c}
{\boldsymbol \alpha}\times{\boldsymbol \beta} \equiv  \pm a b{\bf
e}_3 , \quad {\boldsymbol \omega } =
\varphi ^ {\boldsymbol \cdot}  {\bf e}_3 , \\
{\boldsymbol \alpha } = a({\bf e}_1 \cos \varphi  - {\bf e}_2 \sin
\varphi ),\quad {\boldsymbol \beta } =  \pm b({\bf e}_1 \sin \varphi
+ {\bf e}_2
\cos \varphi ), \\
\varphi ^{ {\boldsymbol \cdot}  {\boldsymbol \cdot} }  =  - (a \pm
b)\sin \varphi.
\end{array} \label{q2_17}
\end{gather}
\end{lemma}

The proof is obvious. Note that in the case considered the pointed
out families are the only motions with constant direction of the
angular velocity. In particular, the body in two independent
constant fields does not have any uniform rotations.

\section{Critical set of the Kowalevski top in two constant fields}\label{sec3}
Suppose that the irreducible RBD-problem has a diagonal inertia tensor
with principal moments of inertia satisfying the ratio 2:2:1; then
we obtain the integrable case \cite{b2} of the Kowalevski top in two
constant fields. By an appropriate choice of measurement units, we present
equations (\ref{q2_6}) in scalar form
\begin{equation}\label{q3_1}
\begin{array}{c}
2\omega _1^ {\boldsymbol \cdot}   = \omega _2 \omega _3  + \beta _3
,\; 2\omega _2^ {\boldsymbol \cdot}   =  - \omega _1 \omega _3  -
\alpha _3 ,\;
\omega _3^ {\boldsymbol \cdot}   = \alpha _2  - \beta _1 , \\
\alpha _1^ {\boldsymbol \cdot}   = \alpha _2 \omega _3  - \alpha_3
\omega_2,\; \beta _1^ {\boldsymbol \cdot}   = \beta _2 \omega _3  -
\beta_3 \omega_2, \\
\alpha _2^ {\boldsymbol \cdot}   = \alpha _3 \omega _1  - \alpha_1
\omega_3,\; \beta _2^ {\boldsymbol \cdot}   = \beta _3 \omega _1  -
\beta_1 \omega_3, \\
\alpha _3^ {\boldsymbol \cdot}   = \alpha _1 \omega _2  - \alpha_2
\omega_1,\; \beta _3^ {\boldsymbol \cdot}   = \beta _1 \omega _2  -
\beta_2 \omega_1.
\end{array}
\end{equation}

The phase space is $P^6={\bf R}^3 \times \mathcal{O}$, where
$\mathcal{O} \subset {\bf R}^3 \times {\bf R}^3$ is defined by
(\ref{q2_11}); $\mathcal{O}$ is diffeomorphic to $SO(3)$.

The complete set of first integrals in involution on $P^6 $ consists
of the energy integral $H$, the generalized Kowalevski integral $K$
\cite{b3}, and the integral $G$ found in \cite{b2}:
\begin{equation}\label{q3_2}
\begin{array}{l}
\displaystyle{H = \omega _1^2  + \omega _2^2  + \frac{1} {2}\omega
_3^2 -(\alpha _1  + \beta _2 ),} \\
K = (\omega _1^2  - \omega _2^2  + \alpha _1  - \beta _2 )^2  +
(2\omega _1 \omega _2  + \alpha _2  + \beta _1 )^2, \\
\displaystyle{G =\frac{1} {4} (2\alpha _1 \omega _1  + 2\alpha _2
\omega _2 +\alpha _3 \omega _3 )^2  + \frac{1} {4} (2\beta _1
\omega _1  +2\beta _2 \omega_2  + \beta _3 \omega _3 )^2  +}  \\
\displaystyle{ \qquad {} + \frac{1} {2} \omega _3 (2\gamma _1 \omega
_1  + 2\gamma _2 \omega _2  + \gamma _3 \omega _3 ) - b^2 \alpha _1
-  a^2 \beta _2.}
\end{array}
\end{equation}
Here we denote by $\gamma _i $ the components of ${\boldsymbol
\gamma}={\boldsymbol \alpha} \times {\boldsymbol \beta}$ relative to
the moving basis.

Introduce the integral map
\begin{equation}\label{q3_3}
J=G\times K\times H: P^6 \rightarrow {\bf R}^3.
\end{equation}

Let $\sigma \subset P^6$ be the set of critical points of $J$. By
definition, the bifurcation diagram of $J$ is the subset $\Sigma
\subset {\bf R}^3$ over which $J$ fails to be locally trivial;
$\Sigma$ determines the cases when the topological type of the integral manifolds
\begin{equation}\label{q3_4}
J_c=J^{-1}(c),   \quad c=(g,k,h) \in {\bf R}^3
\end{equation}
changes. Finding the critical set $\sigma $ and the
bifurcation diagram is the necessary step in the topological analysis of
the problem as a whole.

It follows from the Liouville -- Arnold theorem that for $c \notin
\Sigma $ the manifold (\ref{q3_4}), if not empty, is a union of
three-dimensional tori. The considered Hamiltonian system is
non-degenerate (at least for sufficiently small  values of $b$);
then the trajectories on such a torus are quasi-periodic with three
almost everywhere independent frequencies. The critical set $\sigma$
is invariant under the phase flow and consists of trajectories with
number of frequencies less than three. These trajectories are called
{\it critical motions}. For a generic value $c \in \Sigma $  the set
$J_c \cap \sigma $ consists of two-dimensional tori. The dynamical
system induced on the union of such tori for $c$ in some open subset
in $\Sigma $ is a Hamiltonian system with two degrees of freedom.
Vice versa, let $M$ be a submanifold of $P^6$, $\dim M = 4$, and
suppose that the induced system on $M$ is Hamiltonian. Then,
obviously, $M \subset \sigma$. This speculation gives a useful tool
to find out whether a common level of functions consists of critical
points of $J$.

\begin{lemma}\label{lem5}
Consider a system of equations
\begin{equation}\label{q3_5}
f_1=0,\;f_2=0
\end{equation}
on a domain $W$ open in $P^6$. Let $X$ be the vector field on $P^6$
corresponding to (\ref{q3_1}) and $M \subset W$ defined by
(\ref{q3_5}). Suppose

(i) $f_1$ and $f_2$ are smooth functions independent on $M$;

(ii) $Xf_1=0,\; Xf_2=0$ on $M$;

(iii) the Poisson bracket $\{f_1,f_2\}$ is non-zero almost
everywhere on $M$.

Then $M$ consists of critical points of the map $J$.
\end{lemma}
\begin{proof}
Conditions (i), (ii) imply that $M$ is a smooth four-dimensional
manifold invariant under the restriction of the phase flow to the open
set $W$. Condition (iii) means that the closed \mbox{2-form} induced
on $M$ by the symplectic structure on $P^6$ is almost everywhere
non-degenerate. Thus the flow on $M$ is almost everywhere
Hamiltonian with two degrees of freedom. It inherits the property of
complete integrability. Then almost all its integral manifolds
consist of two-dimensional tori and necessarily lie in $\sigma$.
Since $M$ is closed in $W$ and $\sigma$ is closed in $P^6$, we
conclude that $M \subset \sigma$.
\end{proof}

Two systems of the type (\ref{q3_5}) are known. The first one was
pointed out in \cite{b3}. It is the zero level of the integral $K$.
The condition $K = 0$ leads to two independent equations defining
the smooth four-dimensional manifold $\mathfrak{M} \subset \sigma $.
It is shown in \cite{b10} that the \mbox{2-form} induced on
$\mathfrak{M}$ by the symplectic structure on $P^6$ is degenerate on
the surface of codimension 1.

The second critical subset $\mathfrak{N} \subset \sigma $ was found
in \cite{b11} in the form of a system of two equations satisfying
the conditions of Lemma \ref{lem5}. The functions in these equations
have essential singularities at the points
\begin{equation}\label{q3_6}
\alpha _1  = \beta _2 ,\quad \alpha _2  =  - \beta _1 .
\end{equation}

The set $\mathfrak{N}$ was investigated in \cite{b12}. It was shown
that $\mathfrak{N}$ is the set of critical points of some smooth
function $F$ on $P^6$. Then $\mathfrak{N}$ is stratified by the rank
of Hesse's matrix of $F$ and fails to be a smooth four-dimensional
manifold at some points of the set (\ref{q3_6}). In particular, it
cannot be defined by any global system of two independent equations.
In this case the induced \mbox{2-form} also has degenerate points
even in the smooth part of $\mathfrak{N}$.

The following result completes the description of the critical set
$\sigma$ by adding a new invariant subset $\mathfrak{O} \subset
P^6$; $\mathfrak{O}$ is almost everywhere a smooth four-dimensional
manifold. Note that the sets $\mathfrak{M}, \mathfrak{N}$ and
$\mathfrak{O}$ have pairwise nonempty intersections corresponding to
bifurcations of critical integral manifolds of the induced "almost
Hamiltonian"\, systems with two degrees of freedom.

Let us introduce the following notation
$$
\begin{array}{c}
p^2  = a^2  + b^2 ,\;r^2  = a^2  - b^2 ;\\
\xi_1 = \alpha_1 - \beta_2,\; \xi_2 = \alpha_2 +\beta _1,\;\eta_1 =
\alpha_1  + \beta_2 ,\;\eta_2 = \alpha_2 - \beta_1.
\end{array}
$$

\begin{theorem} \label{th1} The set of critical points of the integral map
(\ref{q3_3}) consists of the following subsets in $P^6$:

1) the set $\mathfrak{M}$ defined by the system
\begin{equation}\label{q3_7}
Z_1  = 0,\;Z_2  = 0
\end{equation}
with
\begin{equation}\label{q3_8}
Z_1  = \omega _1^2  - \omega _2^2
+ \xi _1 ,\;Z_2  = 2\omega _1 \omega _2  + \xi _2 ;
\end{equation}

2) the set $\mathfrak{N}$ defined by the system
\begin{equation}\label{q3_9}
F_1  = 0,\;F_2  = 0,\quad \xi _1^2  + \xi _2^2  \ne 0
\end{equation}
with
\begin{equation}\label{q3_10}
\begin{array}{l}
F_1  = (\xi _1^2  + \xi _2^2 )\omega _3  - 2[(\xi _1 \omega _1 + \xi
_2 \omega _2 )\alpha _3  + (\xi _2 \omega _1  - \xi _1 \omega
_2 )\beta _3 ],  \\
F_2  = (\xi _1^2  - \xi _2^2 )(2\omega _1 \omega _2  + \xi _2 )- 2
\xi _1 \xi _2 (\omega _1^2  - \omega _2^2  + \xi _1 ),
\end{array}
\end{equation}
and by the system
\begin{equation}\label{q3_11}
\begin{array}{c}
\xi _1  = \xi _2  = 0,\;\alpha _3  =  \pm r,\;\beta _3  = 0,\;\eta
_1^2  + \eta _2^2  =2( p^2  - r^2) , \\
(\omega _1^2  + \omega _2^2 )(\alpha _3 \omega _3  + \eta _1 \omega
_1  + \eta _2 \omega _2 ) + r^2 \omega _1  = 0;
\end{array}
\end{equation}

3) the set $\mathfrak{O}$ defined by the system
\begin{equation}\label{q3_12}
R_1  = 0,\;R_2  = 0
\end{equation}
with
\begin{equation}\label{q3_13}
\begin{array}{l}
R_1  = (\alpha _3 \omega _2  - \beta _3 \omega _1 )\omega _3  + 2\xi
_1 \omega _1 \omega _2  - \xi _2 (\omega _1^2  - \omega _2^2
) + \eta _2 (\omega _1^2  + \omega _2^2 ),  \\
R_2  = (\alpha _3 \omega _1  + \beta _3 \omega _2 )\omega _3^2 +
[\alpha _3^2  + \beta _3^2  + \xi _1 (\omega _1^2  - \omega _2^2 )
+ 2\xi _2 \omega _1 \omega _2  +   \\
\qquad {} + \eta _1 (\omega _1^2  + \omega _2^2 )]\omega _3  + 2[\xi
_1 (\alpha _3 \omega _1  - \beta _3 \omega _2 ) + \xi _2 (\alpha _3
\omega _2  + \beta _3 \omega _1 )].
\end{array}
\end{equation}
\end{theorem}

\begin{proof}
Introduce the change of variables \cite{b11} ($i^2=-1$)
\begin{equation}\label{q3_14}
\begin{array}{c}
x_1=\xi_1+ i \xi_2,\quad x_2=\xi_1 - i \xi_2,\\
y_1=\eta_1+ i \eta_2,\quad y_2=\eta_1 - i \eta_2,\\
z_1  = \alpha_3 + i\beta_3, \quad z_2  = \alpha_3 - i\beta_3,\\
w_1=\omega_1 + i\omega_2, \quad w_2  = \omega_1 - i\omega_2, \quad
w_3 = \omega _3.
\end{array}
\end{equation}

The system (\ref{q3_1}) takes the form
\begin{equation}\label{q3_15}
\begin{array}{c}
\begin{array}{ll}
{x'_1  =  - x_1 w_3  + z_1 w_1,} & {x'_2  = x_2 w_3  - z_2 w_2,} \cr
{y'_1  =  - y_1 w_3  + z_2 w_1,} & {y'_2  = y_2 w_3  - z_1 w_2 ,}
\cr {2z'_1  = x_1 w_2  - y_2 w_1,} & {2z'_2  =  - x_2 w_1 + y_1
w_2,}
\end{array} \\
2w'_1  =  - (w_1 w_3  + z_1 ),\quad 2w'_2  = w_2 w_3  + z_2, \quad
2w'_3 = y_2  - y_1.
\end{array}
\end{equation}
Here the prime stands for $d/d(it)$.

Denote by $V^9$ the subspace of ${\bf C}^9$ defined by
(\ref{q3_14}). On $V^9$, equations (\ref {q2_11}) of the phase space
$P^6$ become
\begin{equation}\label{q3_16}
\begin{array}{c}
z_1^2  + x_1 y_2  = r^2 ,\quad z_2^2  + x_2 y_1  = r^2 , \\
x_1 x_2  + y_1 y_2  + 2z_1 z_2  = 2p^2 .
\end{array}
\end{equation}

By virtue of  (\ref{q3_14}) and (\ref{q3_16}) the integrals (\ref {q3_2}) take the form
\begin{equation}\label{q3_17}
\begin{array}{l}
\displaystyle{ H = \frac{1}{2}w_3^2 + w_1 w_2  - \frac{1}{2}(y_1 +
y_2 )
},  \\
\displaystyle{
K=(w_1^2 + x_1 )(w_2^2  + x_2 )},  \\
\displaystyle{ G = \frac{1}{4}(p^2  - x_1 x_2 )w_3^2
+\frac{1}{2}(x_2 z_1 w_1 +x_1 z_2 w_2 )w_3  +}  \\
\displaystyle{\qquad{} + \frac{1}{4}(x_2 w_1  + y_1 w_2 )(y_2 w_1 +
x_1 w_2 ) - \frac{1}{4}p^2 (y_1  + y_2 ) +\frac{1} {4}r^2 (x_1 + x_2
)}.
\end{array}
\end{equation}

Let $f$ be an arbitrary function on $V^9$. For brevity, the term
"critical point of $f$"\, will always mean a critical point of the
restriction of $f$ to $P^6$. Similarly, $df$ means the restriction
of the differential of $f$ to the set of vectors tangent to $P^6$.

While calculating critical points of various functions (in the above
sense), it is convenient to avoid introducing Lagrange multipliers
for the restrictions (\ref{q3_16}). Notice that the following vector
fields
\begin{equation}
\begin{array}{l}
\displaystyle{X_1=\frac{\partial}{\partial w_1},
\;X_2=\frac{\partial}{\partial
w_2},\;X_3=\frac{\partial}{\partial w_3}, }\\
\displaystyle{ Y_1= z_2 \frac{\partial}{\partial x_2} + z_1
\frac{\partial}{\partial y_2} -\frac{1}{2}x_1 \frac{\partial}
{\partial z_1} - \frac{1}{2}y_1\frac{\partial}{\partial z_2}},\\
\displaystyle{ Y_2= z_1 \frac{\partial}{\partial x_1} + z_2
\frac{\partial}{\partial y_1}  - \frac{1}{2} y_2 \frac
{\partial }{ \partial z_1}-\frac{1}{2}x_2\frac{\partial}{\partial z_2}},\\
\displaystyle{ Y_3= x_1 \frac{\partial}{\partial x_1} - x_2
\frac{\partial}{\partial x_2} + y_1 \frac{\partial}{\partial y_1} -
y_2 \frac{\partial}{\partial y_2 }}
\end{array} \notag
\end{equation}
are tangent to $P^6 \subset V^9$ and linearly independent at any
point of $P^6$. Then the set of critical points of $f$ is defined by
the system of equations
\begin{eqnarray}
X_1 f=0, & X_2 f=0, & X_3 f=0, \label{q3_27} \\
Y_1 f=0, & Y_2 f=0, & Y_3 f=0. \label{q3_28}
\end{eqnarray}

1. Apply (\ref{q3_27}) and (\ref{q3_28}) to $f=K$. Then a critical
point of $K$ satisfies either
\begin{equation}\label{q3_18}
w_1^2  + x_1  = 0,\;w_2^2  + x_2  = 0
\end{equation}
or
\begin{equation}\label{q3_19}
w_1  = w_2  = 0,\;z_1  = z_2  = 0.
\end{equation}

The system (\ref{q3_18}) coincides with (\ref{q3_7}) and the only
invariant set generated by (\ref{q3_19}) consists of all points of
the trajectories (\ref{q2_17}). Such points satisfy (\ref{q3_12}).

2. Consider the regular points of $K$ at which $H$ and $K$ are
dependent. Applying (\ref{q3_27}) to $f=H+sK$ with Lagrange
multiplier $s$ we immediately obtain $w_3=0$. Then from (\ref{q3_15})
we come to solutions (\ref{q2_15}), (\ref{q2_16}). Along the
corresponding trajectories both conditions (\ref{q3_9}),
(\ref{q3_12}) are valid.

3. We now assume that $H$ and $K$ are independent. Introduce the
function with Lagrange multipliers $\tau ,s$
\begin{equation}\label{q3_20} \notag
L = 2G + (\tau  - p^2 )H + sK.
\end{equation}
The multiplier of $G$ is non-zero by assumption. The term with $p^2
$ is added for convenience.

The set $\sigma _0  \subset \sigma $ of the points satisfying for
some $\tau ,s$ the condition
\begin{equation}\label{q3_21}
2dG + (\tau  - p^2 )dH + sdK = 0
\end{equation}
is preserved by the phase flow of (\ref {q3_15}). Applying the
corresponding Lie derivative to (\ref{q3_21}) gives
$$
\tau 'dH + s'dK = 0.
$$

Since $dH$ and $dK$ are supposed to be linearly independent, on $\sigma _0 $ we obtain
\begin{equation}\label{q3_22}
\tau ' = 0,\;s' = 0.
\end{equation}

Hence $\tau ,s$ are partial integrals of motion on the invariant
surface $\sigma _0 $.

Equations (\ref{q3_27}) with $f=L$ give
\begin{gather}
\begin{array}{l}
x_2 z_1 w_3  + x_2 y_2 w_1  + (\tau  - z_1 z_2 )w_2  + 2sw_1
(w_2^2  + x_2 ) = 0,  \\
x_1 z_2 w_3  + (\tau  - z_1 z_2 )w_1  + x_1 y_1 w_2  + 2sw_2 (w_1^2
+ x_1 ) = 0;
\end{array} \label{q3_23} \\
(\tau  - x_1 x_2 )w_3  + x_2 z_1 w_1  + x_1 z_2 w_2  = 0.
\label{q3_24}
\end{gather}

First consider the case (\ref{q3_6}). From (\ref{q3_16}) we come to
the following values
\begin{equation}\label{q3_25}
x_1  = x_2  = 0,\;z_1^2  = z_2^2  = r^2 ,\;y_1 y_2  = 2(p^2  - r^2).
\end{equation}

Equations (\ref{q3_23}) and  (\ref{q3_24}) hold if $w_1 w_2 = 0$ or $w_3
= 0$. If either of these equalities takes place on some interval of
time (and hence identically), then we obtain one of the solutions
(\ref{q2_15}) -- (\ref{q2_17}).

Let $w_1 w_2  \ne 0,w_3 \ne 0$ at some point satisfying
(\ref{q3_25}). Then (\ref{q3_23}) and (\ref {q3_24}) yield
\begin{equation}\label{q3_26}
\tau  = 0,\;s = r^2 /(2w_1 w_2 ).
\end{equation}

Since $z_1$ and $z_2$ are complex conjugates of each other, it follows
from (\ref{q3_25}) that they are real and equal. Denote their value by $z = \pm r$.

With (\ref{q3_25}) and (\ref{q3_26}) the system (\ref{q3_28})
reduces to a single equation
\begin{equation}\label{q3_29}
w_1 w_2 [2zw_3  + (w_2 y_1  + w_1 y_2 )] + r^2 (w_1  + w_2 ) = 0,
\end{equation}
which corresponds to (\ref{q3_11}).

Note that (\ref{q3_29}) is obtained from (\ref{q3_9}) as $\rho =
\displaystyle{\sqrt {\xi _1^2 + \xi _2^2 }}$ tends to zero only
after dividing by the maximal available power of $\rho$. Thus, at the points (\ref{q3_6}) the
system (\ref{q3_9}), without the assumption that $\rho \ne 0$, has
extra solutions not belonging to $\sigma$.

Suppose $x_1 x_2  \ne 0$. The determinant of (\ref{q3_23}) with
respect to $\tau ,s$ equals $\delta = 2(x_1 w_2^2  - x_2 w_1^2 )$.
Let $\delta \equiv 0$ on some interval of time; calculating the
derivatives of this identity in virtue of (\ref{q3_15}), we obtain
one of the cases (\ref{q3_18}), (\ref{q3_19}). Therefore we may
assume that $\delta \ne 0$. Then (\ref{q3_23}) implies
\begin{align}
s &= \displaystyle {\frac{1} {{2(x_1 w_2^2  - x_2 w_1^2 )}}}[(x_2
z_1 w_1  -
x_1 z_2 w_2 )w_3  + x_2 y_2 w_1^2  - x_1 y_1 w_2^2 ], \label{q3_31} \\
\tau &= z_1 z_2  + \displaystyle {\frac{1} {{x_1 w_2^2  - x_2 w_1^2
}}}\{ [x_1 x_2 (z_2 w_1  - z_1 w_2 ) - w_1 w_2 (x_2 z_1 w_1
-x_1 z_2 w_2 )]w_3-  \label{q3_32}\\
& \quad {} - w_1 w_2 (x_2 y_2 w_1^2  - x_1 y_1 w_2^2 ) + x_1 x_2 w_1
w_2 (y_1  - y_2 )\} . \notag
\end{align}

Eliminating $\tau $ from (\ref{q3_24}) and (\ref{q3_32}) we obtain $
S_1  = 0$, where
$$
\begin{array}{l}
S_1  = [x_1 x_2 (z_2 w_1  - z_1 w_2 ) - w_1 w_2 (x_2 z_1 w_1  -
x_1 z_2 w_2 )]w_3^2  +  \\
\qquad {} + [(x_1 x_2  - z_1 z_2 )(x_2 w_1^2  - x_1 w_2^2 ) - w_1
w_2 (x_2 y_2 w_1^2  - x_1 y_1 w_2^2 ) +  \\
\qquad {}  + x_1 x_2 w_1 w_2 (y_1  - y_2 )]w_3  - (x_2 w_1^2  - x_1
w_2^2 )(x_2 z_1 w_1  + x_1 z_2 w_2 ).
\end{array}
$$

Next we solve  (\ref{q3_24}) for $\tau $ and calculate the derivative
$\tau'$ in virtue of (\ref{q3_15}).  According to (\ref{q3_22}) we
must have $ S_2  = 0$, where
$$
\begin{array}{l}
S_2  = (x_2 z_1 w_1  - x_1 z_2 w_2 )w_3^2  + (x_2 y_2 w_1^2  - x_1
y_1 w_2^2  + x_2 z_1^2  - x_1 z_2^2 )w_3  -  \\
\qquad {}   - (y_1  - y_2 )(x_2 z_1 w_1  + x_1 z_2 w_2 ).
\end{array}
$$

Notice that
$$
S_1  + w_1 w_2 S_2  = F_1 R.
$$

Here
\begin{equation}\label{q3_32a}
F_1  = x_1 x_2 w_3  - (x_2 z_1 w_1  + x_1 z_2 w_2 )
\end{equation}
is  the first function from (\ref{q3_10}). The
function
\begin{equation}\label{q3_33}
R  = (z_2 w_1  - z_1 w_2 )w_3  + x_2 w_1^2  - x_1 w_2^2  + w_1 w_2
(y_1  - y_2 )
\end{equation}
is a multiple of  the first function from (\ref{q3_13}), precisely,
$R=2iR_1$. Thus on the trajectories consisting of critical points,
we have either $F_1 \equiv 0$ or $R_1 \equiv 0$. Calculating the
derivatives of these identities in virtue of (\ref{q3_15}) we obtain
 (\ref{q3_9}) and (\ref{q3_12}), respectively. Hence (\ref{q3_9}) and
(\ref{q3_12}) provide  necessary conditions for a point to belong
to $\sigma _0 $.

To prove sufficiency, it is enough to check (\ref{q3_28}). We avoid
this technically complicated procedure and only notice that the
systems (\ref{q3_9}) and (\ref{q3_12}) satisfy the assumptions of
Lemma~\ref{lem5}.
\end{proof}

The phase topology of the induced system on $\mathfrak{M}$ was
studied in \cite{b10}. The system of invariant relations
(\ref{q3_9}) corresponds to that found in \cite{b11}. In the paper
\cite{b12} the equations of motion on $\mathfrak{N}$ are separated
and the initial phase variables are expressed via two auxiliary
variables, the latter being elliptic functions of time. The motions
on $\mathfrak{M}$ generalize those of the 1st Appelrot class (Delone
class) of the Kowalevski problem \cite{b13}. As $b$ tends to zero
the motions on $\mathfrak{N}$, as shown in \cite{b11}, convert to
the so-called {\it especially marvelous} motions of the 2nd and 3rd
classes of Appelrot \cite{b13}. The set defined by the system
(\ref{q3_12}) was not pointed out earlier.

To find the classical analog of the set $\mathfrak{O}$, put
${\boldsymbol \beta } = 0$ in (\ref{q3_13}). Then $\xi _1 = \eta _1
= \alpha _1$, $\xi _2  = \eta _2  = \alpha _2 $ and we obtain
$$
R_1  = 2\ell \omega _2 ,\quad R_2  = 2\ell (\omega _1 \omega _3  +
\alpha _3 ),
$$
where $ 2\ell  = 2\alpha _1 \omega _1  + 2\alpha _2 \omega _2  +
\alpha _3 \omega _3 $ is a constant of the momentum integral
existing in the case of one force field. Therefore equations
(\ref{q3_12}) yield either $\ell = 0$ or
\begin{equation}\label{e37}
\omega _2  = 0,\;\omega _1 \omega _3  + \alpha _3  = 0.
\end{equation}

The condition $\ell  = 0$ follows from the fact that when
${\boldsymbol \beta} = 0$ the integral $G$ takes the value $\ell ^2
$. Equations (\ref {e37}) define {\it especially marvelous} motions
of the 4th class of Appelrot~\cite{b13}.

\section{Bifurcation diagram}\label{sec4}
Since all common levels of the first integrals (\ref{q3_2}) are
compact, the bifurcation diagram $\Sigma$ coincides with the set of
critical values of the map (\ref{q3_3}), that is, $\Sigma = J(\sigma )$.

Let $\gamma  = \left| {{\boldsymbol \alpha } \times {\boldsymbol
\beta }} \right|$. According to (\ref{q2_11}), $\gamma  = ab$.

Denote by $\Delta $ the region of existence of motions, that is, the
set of $c=(g,k,h) \in {\bf R}^3$ for which the integral manifolds
(\ref{q3_4}) are not empty.

\begin{theorem} \label{th2}
The bifurcation diagram of the map $G \times K \times H$ is the
intersection of $\Delta$ with the union of the surfaces
\begin{eqnarray}
& & \Gamma _1 :\quad k = 0; \label{q4_1}\\
& & \Gamma _2 :\quad p^2 h- 2 g  + r^2 \sqrt{k} = 0; \label{q4_2}\\
& & \Gamma _3 :\quad p^2 h- 2 g  - r^2 \sqrt{k} = 0; \label{q4_2a}\\
& & \Gamma _4 :\quad \left\{ {\begin{array}{l}
\displaystyle{k = 3 s^2 - 4 h s + p^2 + h^2 - \frac{\gamma^2} {s^2}} \\
\displaystyle{g = -s^3 + h s^2 + \frac{\gamma^2}{s}}
\end{array}} \right., \quad s \in {\bf R} \backslash \{0 \} \label{q4_3}
\end{eqnarray}
and the line segment
\begin{equation}\label{q4_3s}
\Gamma _5 :\quad g=\gamma h, \; k=p^2-2\gamma,\; h^2 \leqslant
4\gamma.
\end{equation}

In the parametric representation of the surface $\Gamma _4 $ the parameter
$s$ stands for a multiple root of the polynomial
\begin{equation}\label{q4_4}
\Phi(s) = s^4 - 2hs^3 +(h^2 + p^2 - k)s^2  - 2gs + \gamma ^2.
\end{equation}
\end{theorem}

\begin{proof}
1. The equation of the surface (\ref{q4_1}) follows immediately from
(\ref{q3_7}), (\ref{q3_8}), and the expression of $K$
in~(\ref{q3_2}).

2. Relations (\ref{q4_2}), (\ref{q4_2a}) are equivalent to
\begin{equation}\label{q4_4a}
(p^2 h - 2g)^2  - r^4 k=0.
\end{equation}

Introduce the function
$$
F = (p^2 H - 2G)^2  - r^4 K.
$$

For $x_1 x_2 \ne 0$ denote
\begin{equation}\label{q4_6}
U_1  = \sqrt {\frac{{x_2 }} {{x_1 }}(w_1^2  + x_1 )} ,\quad U_2  =
\sqrt {\frac{{x_1 }} {{x_2 }}(w_2^2  + x_2 )} \qquad
(\displaystyle{U_2=\overline{U_1}}).
\end{equation}

From representations (\ref {q3_17}) and (\ref{q3_32a}) we obtain
\begin{equation}\label{q4_5}
\begin{array}{l}
\displaystyle{p^2 H - 2G + r^2 \sqrt K  = \frac{1}{2 x_1
x_2}F_1^2+2{r^2
}(\mathop{\rm Im}\nolimits U_1 )^2,}  \\
\displaystyle{p^2 H - 2G - r^2 \sqrt K = \frac{1}{2 x_1
x_2}F_1^2-2{r^2 }(\mathop{\rm Re}\nolimits U_1 )^2.}
\end{array}
\end{equation}
Here $\displaystyle{\sqrt K}$ is the principal square root of $K$.

The equation of the zero level of $F$ splits into two distinct equations
\begin{eqnarray}
F_1^2  + 4 r^2 x_1 x_2 (\mathop{\rm Im}\nolimits U_1 )^2 = 0,\label{q4_7} \\
F_1^2  - 4 r^2 x_1 x_2 (\mathop{\rm Re}\nolimits U_1 )^2 =
0.\label{q4_8}
\end{eqnarray}

From (\ref{q3_10}) and (\ref {q4_6}) we have
$$
F_2  = \frac {x_1 x_2}{2 i}(U_1^2  - U_2^2) = 2 x_1 x_2 \mathop{\rm
Im}\nolimits U_1 \mathop{\rm Re}\nolimits U_1.
$$

Thus the solutions of (\ref{q3_9}) satisfy either (\ref{q4_7}) or
(\ref{q4_8}) and therefore lie on the zero level of the function
$F$. The corresponding values of the first integrals
satisfy~(\ref{q4_4a}).

From (\ref{q3_17}) it follows that (\ref{q4_4a}) holds for all
points of the phase space such that $x_1 x_2  = 0$ (regardless of
their critical or regular nature). Hence (\ref{q4_4a}) holds for the
points~(\ref{q3_11}).

3. Consider the system (\ref{q3_12}). In terms of the variables (\ref{q3_14}) it
is equivalent to the following equations:

\begin{equation}\label {q4_9}
R=0,\quad R_*=0.
\end{equation}
Here $R$ is defined by (\ref{q3_33}) and
\begin{equation}\label {q4_10}
\begin{array}{l}
R_*=(z_2 w_1 +z_1 w_2)w_3^2+[x_2 w_1^2+x_1 w_2^2+w_1 w_2
(y_1+y_2)+2 z_1 z_2]w_3+\\
\qquad {}+2(x_2 z_1 w_1+ x_1 z_2 w_2).

\end{array}
\end{equation}

Notice that, after several differentiations in virtue of the system
(\ref{q3_15}),  the possibility $z_2^2 w_1^2-z_1^2 w_2^2 \equiv 0$
leads to the conditions (\ref{q3_19}), that is, to the critical
motions (\ref{q2_17}). Assuming (\ref{q3_19}) we obtain
from~(\ref{q3_16})
$$
\begin{array}{c}
x_1 x_2=p^2 -2 q,\quad y_1 y_2=p^2+2 q,\quad  (x_1+x_2)y_1
y_2=r^2(y_1+y_2) \\
(q=\pm \gamma).
\end{array}
$$

The corresponding values of the integrals (\ref{q3_17}) are
\begin{equation}\label {q4_10_1}
\displaystyle{ h = \frac{1}{2}w_3^2 - \frac{1}{2}(y_1 + y_2)}
\geqslant -\sqrt{p^2+2q}, \quad k=p^2-2 q, \quad g = q h.
\end{equation}

If $q=-\gamma$, then all of these values satisfy (\ref{q4_3}) with
\begin{equation}\label {q4_10_2}
s=\frac{1}{2}[h \pm \sqrt{h^2+4\gamma}].
\end{equation}

Let $q=\gamma$. Then the values (\ref{q4_10_1}) satisfy (\ref{q4_3})
with
\begin{equation}\label {q4_10_3}
s=\frac{1}{2}[h \pm \sqrt{h^2-4\gamma}],
\end{equation}
that is, only for the energy range $h^2 \geqslant 4\gamma$. For
$q=\gamma$ and $h^2 \leqslant 4\gamma$ the values (\ref{q4_10_1})
fill the segment (\ref{q4_3s}).

Consider the trajectories for which the equalities (\ref{q3_19}) do
not hold identically. Express $w_3$ from the first
equation~(\ref{q4_9}):
\begin{equation} \label{q4_11}
\displaystyle {w_3 = -\frac{1}{z_2 w_1 - z_1 w_2}[x_2 w_1^2  - x_1
w_2^2  + w_1 w_2 (y_1  - y_2 )]}.
\end{equation}

Replacing $w_3$ in $(z_2 w_1 - z_1 w_2)^2 R_*$ by (\ref{q4_11}), we
obtain the expression $ 2 w_1 w_2 Q $ (the resultant of
(\ref{q3_33}), (\ref{q4_10}) as polynomials in $w_3$), where $Q$ is
a non-homogeneous polynomial of third degree in $w_1$ and $w_2$ whose
coefficients are polynomials in $x_i,y_i$ and $z_i$ of degree not
greater than four. Since (\ref{q3_19}) is already excluded, the
system (\ref{q3_12}) is replaced by (\ref{q4_11}) and the equation
\begin{equation} \label{q4_12}
Q =0.
\end{equation}

We claim that in virtue of (\ref{q4_11}) and (\ref{q4_12}), the
values (\ref{q3_31}) and (\ref{q3_32}) satisfy the identities
\begin{equation}\label{q4_13}
\begin{array}{l}
\tau  - p^2  - 2 s(s - H) = 0, \\
(\tau  - p^2 )^2  + 4(p^2  - K)s^2  - 8 G s + (p^4  - r^4 ) = 0, \\
(\tau  - p^2 )(2 s - H) + 2(p^2  - K)s - 2 G = 0.
\end{array}
\end{equation}

Here the calculation sequence is as follows.

We substitute (\ref{q3_17}), (\ref{q3_31}), (\ref{q3_32}),
(\ref{q4_11}) in the left-hand side of each equation (\ref{q4_13})
and multiply the result by the denominator, which is already
supposed to be non-zero. The expression thereby obtained appears to
be the product of some polynomial in variables (\ref{q3_14}) and the
polynomial $Q$, which equals zero due to~(\ref{q4_12}).

Replace in (\ref{q4_13}) the functions $G,K,H$ by their constant
values $g,k,h$ and exclude $\tau $ with the help of the first
relation. The remaining two reduce to the form
\begin{equation}\label{q4_14}
\Phi (s) = 0,\quad d\Phi (s)/ds = 0,
\end{equation}
where $\Phi $ is the polynomial (\ref{q4_4}). Equations (\ref{q4_3})
are equivalent to (\ref{q4_14}).
\end{proof}

{\bf Remark.} It is easy to see now that the relations (\ref{q4_1})
-- (\ref{q4_2a}) turn into corresponding relations of the 1st, 2nd,
and 3rd classes of Appelrot as ${\boldsymbol \beta}$ tends to zero.
Simultaneously, the polynomial (\ref{q4_4}) turns to $s \varphi
(s)$, where $\varphi (s)$ is the Euler resolvent of the second
polynomial of Kowalevski. This provides an alternative insight into the
connection of the set $\Gamma_4$ with the 4th Appelrot class of
motions. The part of the segment $\Gamma_5$ defined by the
inequality $h^2 < 4 \gamma$ for the classical case ($\gamma=0$)
disappears.

\section{The region of existence of motions}\label{sec5}
The results of the previous section are not complete until we find
some conditions that give a criterion to establish whether a point
of $\tilde \Sigma = \Gamma_1 \cup \Gamma_2 \cup \Gamma_3\cup
\Gamma_4\cup \Gamma_5$ belongs to the region of existence of motions
$\Delta=J(P^6) \subset {\bf R}^3$.

Three inequalities of general character can be obtained immediately from
(\ref{q3_2}) and (\ref{q4_5}):
\begin{gather}
k \geqslant 0; \label{q5_1}\\
h \geqslant -(a+b); \label{q5_5}\\
p^2 h \geqslant 2g - r^2 \sqrt {k}.\label{q5_2}
\end{gather}

In case of the Kowalevski top in the gravity field ($p^2=r^2$) the
inequality obtained from (\ref{q5_2}) was established by Appelrot
\cite{b13}.

To get more precise estimations for $(g,k,h) \in \Delta$, restrict
the problem to iso-energetic surfaces $E_h=\{v \in P^6: H(v)=h\}$.
Denote
\begin{equation}\label{q5_3} \notag
J_h = G \times K \left|_{E_h}: E_h \to {\bf R}^2\right.
\end{equation}
and let $\tilde \Sigma_h, \Sigma_h, \Delta_h$ be the cross-sections
of $\tilde \Sigma, \Sigma, \Delta$ by the plane parallel to and height $h$
above the $(g,k)$-plane. For any $h$ the set $\Sigma_h$
is a bifurcation diagram of the map $J_h$.

Notice that all sets $E_h$ are compact. As proved in \cite{b18},
they are  connected as well. Therefore the values of any continuous
function on $E_h$ fill a bounded and connected segment.

Let
\begin{equation}\label{q5_6}
\begin{array}{c}
k_*(h)= \textrm{min}\, K \left| E_h \right.,\; k^*(h)=
\textrm{max}\, K
\left| E_h \right.; \\
g_*(h)= \textrm{min}\, G \left| E_h \right., \; g^*(h)=
\textrm{max}\, G \left| E_h \right. .
\end{array}
\end{equation}

Then the rectangle
\begin{equation}\label{q5_7}\notag
\Pi(h)=\{(k,g): k_*(h) \leqslant k \leqslant k^*(h), \, g_*(h)
\leqslant g \leqslant g^*(h) \}
\end{equation}
cuts $\Sigma_h$ out of $\tilde \Sigma _h$ and we hope that this
operation is not ambiguous.

The following statements allow us to find explicitly the values
(\ref{q5_6}) and give some more information about the sets $\Gamma_i
\cap \Delta$.

We are going to investigate various maps constructed of combinations
of the first integrals $G,K,H$ and, possibly, restricted to
invariant submanifolds in $P^6$. For each map $I: \mathfrak{C} \to
{\bf R}^k$ of this type we call a point $c \in {\bf R}^k$ {\it an
admissible value} if $I^{-1}(c)\neq \emptyset$.

In Sections \ref{sec3}, \ref{sec4} we often referred to the
motions (\ref{q2_15}) -- (\ref{q2_17}). They will be also important in
the sequel. Calculating the related values of $G,K,H$ we obtain the
sets $\lambda_i$ in $(h,g)$-plane and $\mu_i$ in $(h,k)$-plane
($i=1,...,6$):
$$
\begin{array}{lll}
\lambda_1: \; g=a^2h+a(a^2-b^2), & \mu_1: \; k=(h+2a)^2, & h \geqslant -(a+b); \\
\lambda_2: \; g=a^2h-a(a^2-b^2), & \mu_2: \; k=(h-2a)^2, & h \geqslant a-b; \\
\lambda_3: \; g=b^2h-b(a^2-b^2), & \mu_3: \; k=(h+2b)^2, & h \geqslant -(a+b); \\
\lambda_4: \; g=b^2h+b(a^2-b^2), & \mu_4: \; k=(h-2b)^2, & h \geqslant -a+b; \\
\lambda_5: \; g=a b h, & \mu_5: \; k=(a - b)^2, & h \geqslant -(a+b); \\
\lambda_6: \; g=- a b h, & \mu_6: \; k=(a + b)^2, & h \geqslant
-a+b.
\end{array}
$$

The existing pairwise intersections of the first four sets in either
group correspond to the equilibria~(\ref{q2_14}).

Recall that $\mathfrak{M}=\{K=0\} \subset P^6$ and $J(\mathfrak{M})=
\Gamma_1 \cap \Delta$. Denote
\begin{equation}\label{q5_8}\notag
M=p^2H-2G: P^6 \rightarrow {\bf R}
\end{equation}
and let $H^{(1)}=H |_\mathfrak{M},\; M^{(1)}=M |_\mathfrak{M}$. The
following result belongs to D.B.~Zotev \cite{b10}.

\begin{proposition}\label{prop2} (i) The function $H^{(1)}$ has three critical values
$h_1=-2b$, $h_2=2b$ and $h_3=2a$. In particular,
\begin{equation}\label{q5_9}\notag
\min_{\mathfrak{M}} H=-2b.
\end{equation}

(ii) The bifurcation diagram of $J^{(1)}=H^{(1)}\times
M^{(1)}:\,\mathfrak{M} \to {\bf R}^2$ consists of the half-line
\begin{equation}\label{q5_10}
m=0,\; h \geqslant -2b
\end{equation}
and the set of solutions of the equation
\begin{equation}\label{q5_11}
\begin{array}{c}
27 m^4 + 4h(h^2-18p^2)m^3-2[4p^2h^4-
(16p^4+15r^4)h^2+2p^2(8p^4-9r^4)] m^2+\\
+4r^4 h[h^4-4p^2h^2+2(2p^4-3r^4)]m-r^8[(h^2-2p^2)^2-4r^4]=0
\end{array}
\end{equation}
in the quadrant $\{ m \geqslant 0, h \geqslant -2b\}$.

(iii) The set of admissible values of $J^{(1)}$ is
\begin{equation}\label{q5_12}\notag
0 \leqslant m \leqslant m_0(h), \;h \geqslant -2b,
\end{equation}
where $m_0(h)$ stands for the greatest
positive root of (\ref{q5_11}), which is considered as an equation in $m$.
\end{proposition}

The bifurcation diagram of $J^{(1)}$ is shown in Fig.~1. The
admissible values fill the shaded region.

\begin{figure}[ht]\label{fig1}
\centering
\includegraphics[width=8cm,keepaspectratio]{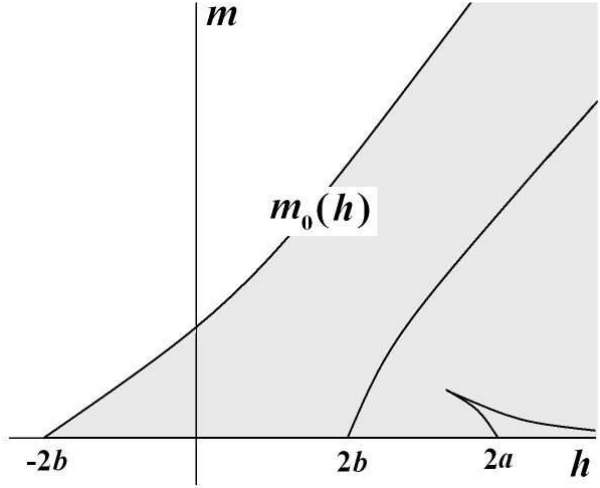}
\caption{The bifurcation diagram of $H^{(1)}\times M^{(1)}$}
\end{figure}

The proof given in \cite{b10} is based on an ingenious change of
variables on $\mathfrak{M}$. Let us point out the relation between
this result and Theorem~\ref{th2}.

Let
\begin{equation}\label{q5_13}
m=p^2h-2g.
\end{equation}

It follows from (\ref{q5_2}) that $m \geqslant 0$ on $\mathfrak{M}$.
This inequality explains (\ref{q5_10}). Moreover, the line $m=0$ in
the plane $k=0$ is the intersection $\Gamma_1 \cap (\Gamma_2\cup
\Gamma_3)$ (in fact, along this line $\Gamma_1$ and $\Gamma_2\cup
\Gamma_3$ are tangent to each other).

The intersection $\Gamma_1 \cap \Gamma_4$ is defined by the system
obtained from (\ref{q4_1}), (\ref{q4_3})
\begin{equation}\label{q5_15}
\begin{array}{l} 3s^4-4h s^3+(p^2+h^2)s^2-\gamma^2=0,\\
s^4-h s^3+g s - \gamma^2=0.
\end{array}
\end{equation}

By virtue of the notation (\ref{q5_13}) the left-hand side of (\ref{q5_11}) becomes
the resultant of the polynomials in (\ref{q5_15}) with respect to
$s$. Thus the set (\ref{q5_11}) corresponds to $\Gamma_1 \cap
\Gamma_4$.

Recall that $\mathfrak{N}\subset P^6$ is defined by (\ref{q3_9}),
(\ref{q3_11}) and $J(\mathfrak{N}) = (\Gamma_2\cup\Gamma_3) \cap
\Delta$. Let $H^{(2)}=H|_\mathfrak{N}$, $G^{(2)}=G|_\mathfrak{N}$.
Introduce the map
\begin{equation}\label{q5_16}\notag
J^{(2)}=H^{(2)}\times G^{(2)}:\,\mathfrak{N} \to {\bf R}^2.
\end{equation}

The following statement is proved in \cite{b14}.

\begin{proposition}\label{prop3} (i) The bifurcation diagram of $J^{(2)}$
consists of the half-lines $\lambda_1$, $\lambda_2$, $\lambda_3$,
$\lambda_4$, the half-line
\begin{equation}\label{q5_17}\notag
g=\frac{1}{2} p^2 h,\; h\geqslant -2b,
\end{equation}
and the curve
\begin{equation}\label{q5_18}\notag
2p^2(p^2 h-2g)^2-2 r^4
h(p^2h-2g)+r^8=0,\;p^2h\geqslant 2g.
\end{equation}

(ii) The admissible values of $J^{(2)}$ fill the region defined by
the system of inequalities
\begin{equation}\label{q5_19}\notag
\left\{ \begin{array}{l} b^2 h-b(a^2-b^2) \leqslant g
\leqslant a^2 h+a(a^2-b^2),\;
h \geqslant -(a+b), \\
2p^2(p^2 h-2g)^2-2 r^4 h(p^2h-2g)+r^8 \geqslant 0.
\end{array} \right.
\end{equation}
\end{proposition}

The bifurcation diagram of $J^{(2)}$ is shown in Fig.~2. The
admissible values fill the shaded region.

\begin{figure}[ht] \label{fig2}
\centering
\includegraphics[width=8cm,keepaspectratio]{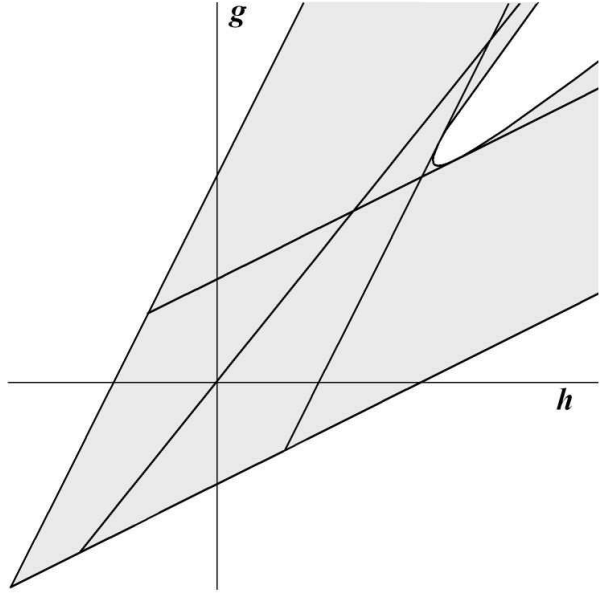}
\caption{The bifurcation diagram of $H^{(2)}\times G^{(2)}$}
\end{figure}

Propositions \ref{prop2}, \ref{prop3} completely define those parts
of $\Gamma_1$, $\Gamma_2$ and $\Gamma_3$ which correspond to real
critical motions, that is, the sets $\Gamma_1 \cap \Delta$ and
$(\Gamma_2\cup\Gamma_3) \cap \Delta$.

Consider the map
\begin{equation}\label{q5_20}\notag
J^{(3)}=H\times K:\,P^6 \to {\bf R}^2.
\end{equation}

The critical set of $J^{(3)}$ is already found (see steps 1, 2 in
the proof of Theorem \ref{th1}). In addition to the manifold
$\mathfrak{M}$ it contains all pendulum motions (\ref{q2_15}) --
(\ref{q2_17}).

\begin{proposition}\label{prop4}
(i) The bifurcation diagram of the map $J^{(3)}$ consists of the
parabolic curves $\mu_1$, $\mu_3$, $\mu_2$, $\mu_4$, the half-lines
$\mu_5$, $\mu_6$, and the half-line
\begin{equation}\label{q5_21}
k=0,\; h\geqslant -2b.
\end{equation}

(ii) Let
\begin{equation}\label{q5_22s}
\begin{array}{l}
k_*(h)=\left\{\begin{array}{ll} (h+2b)^2, &
-(a+b)\leqslant h \leqslant -2b\\
0, & h \geqslant -2b \end{array} \right., \\
k^*(h)=(h+2a)^2.
\end{array}
\end{equation}

The admissible values of $J^{(3)}$ fill the region
\begin{equation}\label{q5_22}
k_*(h) \leqslant k \leqslant k^*(h),
\; h \geqslant -(a+b).
\end{equation}
\end{proposition}

The inequality in (\ref{q5_21}) follows from Proposition
\ref{prop2}. The relationship $k_*(h)$ in (\ref{q5_22s}) is built
in accordance with (\ref{q5_1}). The range of $h$ in (\ref{q5_22})
is defined by (\ref{q5_5}). The region of admissible values (shaded
in Fig.~3) is found using the mentioned above fact that for
each $h \geqslant -(a+b)$ the image of $E_h$ under $K$ is a bounded
connected segment.

\begin{figure}[ht] \label{fig3}
\centering
\includegraphics[width=8cm,keepaspectratio]{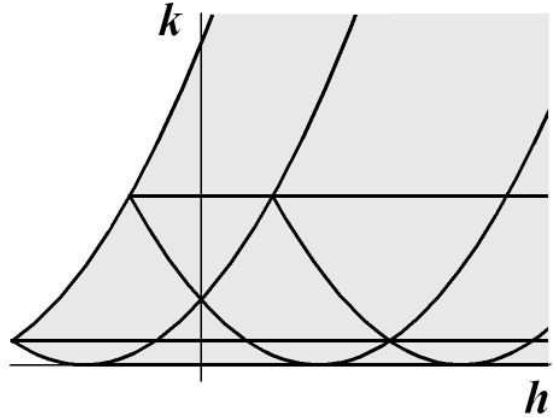}
\caption{The bifurcation diagram of $H \times K$}
\end{figure}

Finally, consider the map
\begin{equation}\label{q5_23}\notag
J^{(4)}=H\times G:\,P^6 \to {\bf R}^2.
\end{equation}

For $ 0 < |s| \leqslant b$ and $|s| \geqslant a$, let
\begin{equation}\label{q5_24}\notag
\phi(s)=\sqrt{\frac{(s^2-a^2)(s^2-b^2)}{s^2}} \geqslant 0.
\end{equation}

\begin{proposition}\label{prop5}
(i) The bifurcation diagram of the map $J^{(4)}$ consists of the
half-lines $\lambda_1$, $\lambda_2$, $\lambda_3$, $\lambda_4$,
$\lambda_5$, $\lambda_6$ and the curves
\begin{eqnarray}
& & C_1: \left\{
\begin{array}{l}
h=2s+\phi(s) \\
\displaystyle{g=\frac{\gamma^2}{s}+s^3+s^2\phi(s)}
\end{array}\right.,\quad s \in [-b,0), \label{q5_25}\\
& & C_2: \left\{
\begin{array}{l}
h=2s+\phi(s) \\
\displaystyle{g=\frac{\gamma^2}{s}+s^3+s^2\phi(s)}
\end{array}\right.,\quad s \in (0,b\,], \label{q5_26}\\
& & C_3: \left\{
\begin{array}{l}
h=2s-\phi(s) \\
\displaystyle{g=\frac{\gamma^2}{s}+s^3-s^2\phi(s)}
\end{array}\right.,\quad s \in [a,+\infty). \label{q5_27}
\end{eqnarray}

(ii) Denote by $g_0(h)$ the one-valued function defined by
(\ref{q5_25}) when $h \geqslant -2b$. Let
\begin{equation}\label{q5_28}
\begin{array}{l}
g_*(h)=\left\{\begin{array}{ll} b^2h-b(a^2-b^2), &
-(a+b)\leqslant h \leqslant -2b\\
g_0(h), & h \geqslant -2b \end{array} \right., \\
g^*(h)=a^2h+a(a^2-b^2).
\end{array}
\end{equation}

Then the region of admissible values of $J^{(4)}$ is defined by the
inequalities
\begin{equation}\label{q5_29}\notag
g_*(h) \leqslant g \leqslant g^*(h),\; h \geqslant -(a+b).
\end{equation}
\end{proposition}

\begin{figure}[ht] \label{fig4}
\centering
\includegraphics[width=8cm,keepaspectratio]{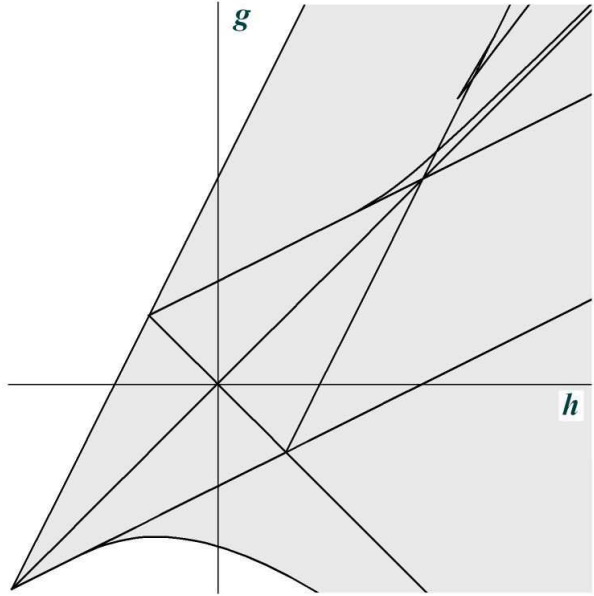}
\caption{The bifurcation diagram of $H\times G$}
\end{figure}

The bifurcation diagram of $J^{(4)}$ with the shaded region of
admissible values is shown in Fig.~4.

A straightforward proof of Proposition~\ref{prop5} can be obtained
using the same technique as in the proof of Theorem~\ref{th1}. Here
we just point out some general ideas that explain this result from
the point of view of the geometry of $\Sigma$.

Let $v \in P^6$ be the critical point of $J^{(4)}$. If $dH(v)=0$,
then $v$ is an equilibrium, that is, a singular point of the system
(\ref{q3_1}). Such a point is a critical point for each first integral
of (\ref{q3_1}). In particular, $dG(v)=0$. The values $(h,g)$ of
$J^{(4)}$ at equilibria (\ref{q2_14}) are the points of pairwise  intersection
of the lines $\lambda_1 - \lambda_4$.

Let
\begin{equation}\label{q5_30}
\mathop{\rm rank}\nolimits \{dG(v),dH(v)\}=1,
\end{equation}
and $c=(g,k,h)=J(v)$. Then $c \in \Sigma$. It follows from
(\ref{q4_1}) -- (\ref{q4_3s}) that (\ref{q5_30}) necessarily implies
\begin{equation}\label{q5_31}
\mathop{\rm rank}\nolimits
\{dG(v),dH(v), dK(v)\}=1.
\end{equation}

If the tangent plane to $\Sigma$ at the point $c$ is well defined,
then the set of zero linear combinations of $dG(v),dH(v)$ and
$dK(v)$ is one-dimensional. This fact contradicts to (\ref{q5_31}).
Therefore, $c$ belongs either to the segment $\Gamma_5$ or to the
set of transversal intersections of two smooth leaves of $\Sigma$.

The intersection of $\Gamma_1$ and  $\Gamma_2\cup \Gamma_3$ is
nowhere transversal.

Transversal intersections of $\Gamma_1$ and $\Gamma_4$ are given by
the system (\ref{q5_15}). Solving it with respect to $g$ and $h$ and
taking into account the admissible region established in
Proposition~\ref{prop2}, we arrive at the curves (\ref{q5_25}) --
(\ref{q5_27}).

Consider intersections of $\Gamma_2\cup\Gamma_3$ and $\Gamma_4$.
Substitute (\ref{q4_3}) for $k,g$ in (\ref{q4_4a}):
\begin{equation}\label{q5_32}
(s^2-a^2)(s^2-b^2)[2s^2-2h s+p^2]^2=0.
\end{equation}

Transversal intersections correspond to the values $s=\pm a,\;s=\pm
b$ (the last multiplier in (\ref{q5_32}) is responsible for the
tangency points of $\Gamma_2\cup\Gamma_3$ and $\Gamma_4$). This
implies the equations of $\lambda_1$ -- $\lambda_4$. As shown in
\cite{b12} the corresponding motions on $\mathfrak{N}$ are the
pendulums (\ref{q2_15}), (\ref{q2_16}). From this fact the
inequalities for $h$ are obtained.

Suppose that $\Gamma_4$ has a point of self-intersection. Then for
some $h$ the curve defined by (\ref{q4_3}) in $(g,k)$-plane has a
double point. Let $s_1,s_2$ be the corresponding values of $s$. It
follows from (\ref{q4_3}) that
$$
s_1+s_2=h,\quad s_1^2 s_2^2=\gamma^2.
$$
Hence $s_1,s_2$ form one of the pairs (\ref{q4_10_2}),
(\ref{q4_10_3}). Substituting these pairs in (\ref{q4_3}) gives
(\ref{q4_10_1}). The obtained set of points in $(h,g)$-plane united
with the projection of the segment $\Gamma_5$ forms the half-lines
$\lambda_5$ and $\lambda_6$.

The admissible region for $(h,g)$ is established in the same way as
in the previous case.

Propositions \ref{prop4} and \ref{prop5} give the explicit formulae
(\ref{q5_22s}), (\ref{q5_28}) for the values (\ref{q5_6}). Then for
each $h$ we can compute the limits for the parameter $s$ in
(\ref{q4_3}) corresponding to $\Gamma_4 \cap \Delta$. Thus the set
$\Sigma_h$ is completely determined. Finally, $\Delta_h$ is obtained
as the span of the curves $\Gamma_i \cap \Delta_h$.

\section{Conclusion}

At this point we can draw all bifurcation diagrams of the induced
momentum maps on iso-energetic surfaces, which are typically
five-dimensional. A lot of information on the stability of the
critical integral manifolds may be immediately obtained for the tori
in $\mathfrak{M}$ and $\mathfrak{N}$. The investigation of the new
critical set $\mathfrak{O}$ waits to be fulfilled.

Since each $E_h$ is a foliation into three-dimensional tori with
some degenerations, we can construct the base $B_h$ for such a
foliation just by factorizing $E_h$, more exactly, by identifying
points of the same connected component of $J_{g,k,h}$. Then $B_h$ is
a two-dimensional analog of Fomenko's graph \cite{b5} for an
iso-energetic manifold of integrable system with two degrees of
freedom. In its turn $B_h$ is a bundle over $\Delta_h$ whose fibres
are finite sets; the number of elements in any fibre is equal to the
number of connected components of the corresponding integral
manifold. The problem of finding this number for all possible
situations seems solvable. Then we obtain a complete description of
the "coverings"\, $B_h \to \Delta_h$ and, consequently, establish
the topology of~$B_h$.

Naturally, the next step requires new mathematical ideas on how the
tori in $E_h$ glue together along the paths in the admissible
regions.

If we consider $B_h$ as a two-dimensional cell complex, then, for
regular levels of energy, \mbox{0-cells} correspond to closed
orbits, a point of each \mbox{1-cell} represents a two-dimensional
torus, and a point of each \mbox{2-cell} represents a
three-dimensional torus. The union of the cells of dimensions 0 and
1 forms a graph, to which the method of marked molecules \cite{b16}
can be applied without any modification. The question is what kind
of a numeric mark should be attached to each two-dimensional cell to
obtain from $B_h$ the complete invariant of Liouville's foliation of
the iso-energetic surface?

Another approach is to consider the set $\Sigma_h^0$ of singular
points of $\Sigma_h$ (self-intersections, tangency points, and
cusps), which is easy to obtain from the above results, and
associate to each $c \in \Sigma_h^0$ the marked loop
molecule~\cite{b17}. In this case, of course, the notion of a mark
should be changed to suit increased dimensions of the tori.

We see that the Kowalevski top in two constant fields provides a
highly non-trivial example of integrable Hamiltonian system and a
complete description of its phase topology is really a challenging
problem.

{\bf Received 09.04.05}

\end{document}